\documentclass[useAMS,usenatbib,usegraphicx]{mn2e}

\def\aap{A\&A}
\def\apjl{ApJ}
\def\apj{ApJ}
\def\apjs{ApJS}
\def\aj{AJ}
\def\mnras{MNRAS}

\def\pasp{PASP}

\title[11-12 Gyr Old WDs]
{11-12 Gyr old White Dwarfs 30 parsecs Away}

\author[M. Kilic et al.]
       {Mukremin Kilic$^{1}\thanks{Email: kilic@ou.edu}$,
       John R. Thorstensen$^2$,
       P. M. Kowalski$^3$,
       and J. Andrews$^4$\\
       $^1$Homer L. Dodge Department of Physics and Astronomy, University of Oklahoma,
       440 W. Brooks St., Norman, OK, 73019, USA\\
       $^2$Department of Physics and Astronomy, Dartmouth College, 6127 Wilder Laboratory, Hanover, NH 03755, USA\\
       $^3$Helmholtz Centre Potsdam - GFZ German Research Centre for Geosciences, Telegrafenberg, 14473 Potsdam, Germany\\
       $^4$Columbia Astrophysics Laboratory, Columbia University, New York, NY 10027, USA
}

\begin{document}

\maketitle

\begin{abstract}

We present a detailed model atmosphere analysis of two of the oldest stars known in the solar neighborhood,
the high proper motion white dwarfs SDSS J110217.48+411315.4 (hereafter J1102) and WD 0346+246 (hereafter WD0346).
We present trigonometric parallax observations of J1102, which places it at a distance of only 33.7 $\pm$ 2.0 pc.
Based on the state of the art model atmospheres, optical, near-, mid-infrared photometry, and distances, we constrain
the temperatures, atmospheric compositions, masses, and ages for both stars. J1102 is an 11 Gyr old (white dwarf plus
main-sequence age), 0.62 $M_{\odot}$ white dwarf with a pure H atmosphere and $T_{\rm eff}=$ 3830 K.
WD0346 is an 11.5 Gyr old, 0.77 $M_{\odot}$ white dwarf with a mixed H/He atmosphere and $T_{\rm eff} = 3650$ K.
Both stars display halo kinematics and their ages agree remarkably well with the ages of the nearest globular clusters,
M4 and NGC 6397. J1102 and WD0346 are the closest examples of the oldest halo stars that we know of.

\end{abstract}

\begin{keywords}
	stars: distances ---
	stars: Population II ---
        stars: atmospheres ---
	stars: individual (SDSS J110217.48+411315.4, WD 0346+246) ---
        white dwarfs
\end{keywords}

\section{INTRODUCTION}

The majority of the stars with an initial mass less than about 8 $M_{\odot}$ end up as white dwarfs (WDs). The short
main-sequence lifetimes of intermediate-mass stars means that they spend almost all of their lives as cooling WDs, radiating away
their residual thermal energy slowly. To first order there is a simple relation between the age and luminosity of a
WD \citep{mestel52}. A typical 0.6 $M_{\odot}$ pure H atmosphere WD cools down to 3800 K in about
10 Gyr \citep{fontaine01}. \citet{winget87} and \citet{liebert88} were the first ones
to use the oldest WDs in the solar neighborhood to constrain the age of the Galactic disk. Further studies based on
nearby high proper motion WDs demonstrate that the oldest disk WDs are about 8 $\pm$ 1.5 Gyr old
\citep[][and references therein]{leggett98,harris06,kilic10b}.

WD cosmochronology also constrains the ages of the oldest halo WDs in the nearest globular clusters or in
the field. \citet{hansen04,hansen07} use 100+ orbit {\em Hubble Space Telescope (HST)} observations of
M4 and NGC 6397 and derive ages of 12.1 (with a 95\% lower limit of 10.3 Gyr) and 11.5 Gyr, respectively.
Another large {\em HST} program on 47 Tucanae has recently been completed, but an age estimate from the WD cooling sequence
is not yet available \citep{kalirai12}. The extension of this method to halo WDs in the field is more problematic
due to the unknown population membership of the high proper motion objects \citep[see][]{oppenheimer01b,bergeron05}. 

Fast moving halo WDs
must exist in the solar neighborhood, though in relatively small numbers compared to the larger population of younger disk
WDs. For example, the Besan\c{c}on Galaxy model predicts 127 disk and 3 halo WDs per square degree within 1 kpc for a Galactic
latitude of 45$^{\circ}$ \citep{robin03}. However, the halo WDs are predicted to be fainter than $V=22$ mag. Hence, deep, wide-field
photometric and astrometric surveys are usually required to identify a significant population of field halo WDs. Such a survey
based on the SDSS + USNO-B \citep{munn04} and the SDSS + Bok telescope proper motions
identified several spectroscopically confirmed halo WD candidates \citep{kilic10a,kilic10b}.
In addition, there are a dozen ultracool WDs that may be thick disk or halo WDs \citep[e.g.,][]{gates04,harris08}. However, trigonometric
parallax measurements are required to confirm their halo membership.
Large scale surveys like the Panoramic Survey Telescope \& Rapid Response System (Pan-STARRS), Palomar Transient Factory, and the
Large Synoptic Survey Telescope will be extremely useful for the identification of large samples of faint halo WDs \citep{tonry12}.

\begin{figure}
\includegraphics[width=1.65in,angle=-90]{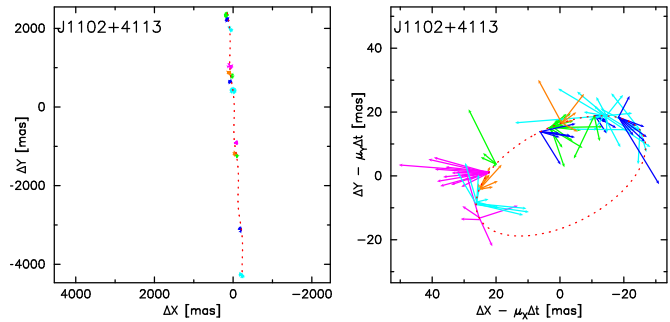}
\caption{The trajectory of J1102 on the
sky. Right panel: The same trajectory with the proper motion taken
out. The tip of each arrow is the position from a single image, and
the tail is the computed location based on the fitted trajectory including
zero point, proper motion, and parallax.}
\end{figure}

Even though the probability of finding a relatively nearby halo WD is small for any small patch of the sky, serendipitious
discoveries of nearby halo WDs have happened. WD0346 and J1102 are the best examples of such discoveries. \citet{hambly97}
identified WD0346 as a high proper motion WD ($1.3\arcsec$ yr$^{-1}$) in the UK Schmidt Telescope photographic plates taken
from 1987 to 1994. Similarly, \citet{hall08} identified J1102 as a high proper motion WD ($1.75\arcsec$ yr$^{-1}$) while searching for high-redshift quasar candidates
in the SDSS spectroscopy data (the SDSS spectrum of J1102 shows a broad emission feature at 5750-5900 \AA, which is an artifact).
\citet{hall08} explain the spectral energy distribution (SED) of J1102 with a 3830 K pure H atmosphere WD. However, without a parallax
measurement, the mass, cooling age, and the kinematic membership of this WD cannot be constrained accurately.

Here we present parallax and mid-infrared photometric observations of J1102,
and a detailed model atmosphere analysis of both stars using parallax, optical, and infrared photometry.
Our observations are discussed in Section 2, whereas our model atmosphere analysis and the nature of these objects are discussed in Section 3 and 4, respectively. 

\section{Observations}

\subsection{Parallax}

We obtained trigonometric parallax observations of J1102 at the MDM 2.4m
Hiltner telescope. We used a total of 106 images taken on 13 observing runs
between 2008 February and 2011 December. The
instrumentation, observing protocols, and reduction techniques used were very
similar to those described in \citet{thor03}, \citet{thor08}, and \citet{kilic08}.
For nearly all observations we used a 2048$^2$ SITe CCD
detector (`Echelle') at the f7.5 focus, but for the last two runs we used a
2048$^2$ STIS CCD (`Nellie'), due to problems with the Echelle CCD. Echelle
has 24 $\mu$m pixels subtending $0.28\arcsec$, and Nellie has 21 $\mu$m pixels
subtending $0.24\arcsec$. At each epoch we took many exposures in a 4-inch
wide Kron-Cousins $I$-band filter, as near to the meridian as we could to
minimize differential color refraction effects. The parallax reduction and
analysis pipeline was unchanged from the previous work, and the change of
detectors appeared to have no effect on the results.

For J1102, we measure a proper motion of $\mu=(-104.8 \pm 0.9, -1741.8 \pm 0.9)$ mas yr$^{-1}$,
and a relative parallax and formal error of $28.8 \pm 1.4$ mas. Using the
colors and magnitudes of the reference-frame stars, and increasing the
uncertainty slightly to allow for unmodeled systematics, we estimate the
absolute parallax to be $29.6 \pm 1.7$ mas. Our proper motion measurement is
relative to the chosen reference stars.
Corrections to absolute proper motions are generally of order
10 mas yr$^{-1}$. Our measurement is consistent with $\mu=(-106.9, -1750.0)$
mas yr$^{-1}$ in the absolute reference frame from \citet{munn04}.

Figure 1 displays the trajectory of J1102 on the sky, and the same trajectory
with proper motion component taken out. This figure shows that our
observations cover a large range of parallax factor for J1102, and the parallax
is well constrained. We use the full Bayesian formalism described in
\citet{thor03} to estimate the distance, including the Lutz-Kelker correction
and prior information from the proper motion and very liberal limits on the
luminosity. J1102 is only 33.7 $\pm$ 2.0 pc away from us.

\subsection{Mid-Infrared Photometry}

We obtained {\em Spitzer} InfraRed Array Camera \citep[IRAC,][]{fazio04} 3.6, 4.5, 5.8, and 7.9 $\mu$m images of WD0346 and J1102 as part of the 
Cycle 3 program 30208 and Cycle 5 program 474. For each object, we obtained 100 second exposures for five dither positions in each filter.
Our reduction procedures are similar to the procedures employed by \citet{K09b}.
Since our targets are relatively faint, we use the smallest aperture (two pixels) for which there are published aperture corrections.
For J1102 we measure $58.1 \pm  2.3, 40.4 \pm  3.0, 35.1 \pm  9.8, 23.3 \pm 13.9~\mu$Jy in channels 1, 2, 3, and 4, respectively.
Unfortunately, WD0346 is blended with a brighter source in the IRAC images, which prohibits us from performing aperture photometry on it. None of these objects are detected in the WISE survey \citep{wright10}.

\section{Model Atmosphere Analysis}

We use the state of the art WD model atmospheres to fit the available photometry for our targets.
The model atmospheres include the Ly-$\alpha$ red wing opacity \citep{kowalski06b} as well as non-ideal
physics of dense helium that includes refraction \citep{kowalski04}, ionization equilibrium \citep{kowalski07},
and the non-ideal dissociation equilibrium of H$_2$ \citep{kowalski06a}.
We convert the magnitudes into monochromatic fluxes using the zero points derived from
the Vega (STIS) spectrum integrated over the passband for each filter. 
We perform a two dimensional least squares fit in $T_{\rm eff}$ and the He/H ratio by minimizing the
differences between the synthetic and the observed fluxes weighted by the observational errors. We use
the parallax measurements to constrain the surface gravity.

\subsection{J1102+4113}

We use the optical and near-infrared photometry \citep{hall08} plus our {\em Spitzer} photometry and distance measurement
to model the J1102 SED. Figure 2 presents our best-fit model compared to the observations.
J1102 is best explained by a $T_{\rm eff}=3830$ K, $\log{g}=8.08$, pure hydrogen atmosphere WD with $M=0.62 M_{\odot}$,
which confirms the temperature assignment of \citet{hall08}. The cooling age of the WD
is $10.0^{+0.4}_{-1.1}$ Gyr. The error in the temperature estimate is about 200 K.
The best-fit model matches the infrared data fairly well; even though the H/He ratio is left as a free parameter in our
model fits, the best-fit model always has a pure hydrogen atmosphere composition for this star. The fit to the optical portion of the SED
is reasonable, but not perfect, possibly due to the problems with the Ly-$\alpha$ opacity calculations. 
The Ly-$\alpha$ opacity calculations from \citet{RAK11} do not result in any significant differences in our model calculations
for this star. The optical portion of the SED favors a slightly cooler solution, whereas the infrared SED favors a hotter solution.
There are known problems with the H$_2$-H$_2$ and H$_2$-He collision-induced absorption \citep[CIA,][]{frommhold10} calculations in dense WD atmospheres.
Even though the strong CIA feature around 2$\mu$m is reproduced in our models, the predicted feature around 1$\mu$m has never been
observed in cool WDs. Regardless of these problems, our best-fit model matches the overall SED fairly well. J1102 is clearly a very cool
and old WD with a pure hydrogen atmosphere.

\begin{figure}
\includegraphics[width=3.2in,angle=0]{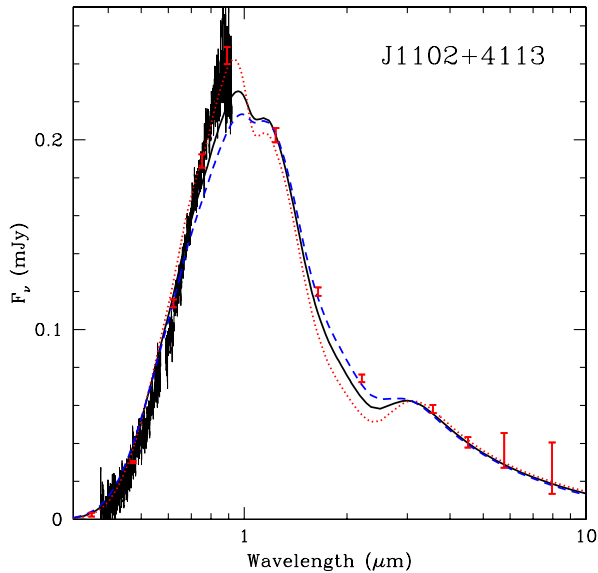}
\caption{The J1102 SED compared to the best-fit pure hydrogen atmosphere model spectrum ($T_{\rm eff}=3830$ K, solid line). Two other models
with $T_{\rm eff}=3600$ K (dotted line) and 4000 K (dashed line) are shown for comparison. The optical spectrum is from the SDSS, with the artifact at 5750-5900 \AA\ blocked out.}
\end{figure}

\subsection{WD 0346+246}

\citet{hambly99} measure a parallax of 36 $\pm$ 5 mas for WD0346, placing it at 28 $\pm$ 4 pc.
The optical to near-infrared SED of WD0346 is kindly made available to us by B. Oppenheimer and
is presented in Figure 3. Using rather crude models for mixed H/He atmosphere WDs, \citet{oppenheimer01a} find a best-fit solution of
$T_{\rm eff}=3750$ K and a He-dominated atmosphere with trace amounts of hydrogen ($\log \rm He/H=9.1$). 
\citet{bergeron01} points out that an extremely helium-rich composition is unlikely for such an old WD, because of accretion
from the interstellar medium. In order to reproduce the short wavelength end of the cool WD spectra, including WD0346,
\citet{bergeron01} adds a pseudocontinuum opacity (due to the bound-free opacity associated with the so-called dissolved atomic
levels of the hydrogen atom). The spectrum of WD0346 is then fit fairly well with a $T_{\rm eff}=3780$ K, $\log{g} =$ 8.34, and He/H =
1.3 model. More recently, \citet{kowalski06b} and \citet{kowalski06c} identified the missing opacity in the blue as the red
wing of the Ly $\alpha$ absorption, and models including this opacity successfully reproduce the short-wavelength fluxes of many cool
WDs \citep{kowalski06b,K08,K09a,K09b,kilic10a,DUR12,giammichele12}.

\begin{figure}
\includegraphics[width=3.2in,angle=0]{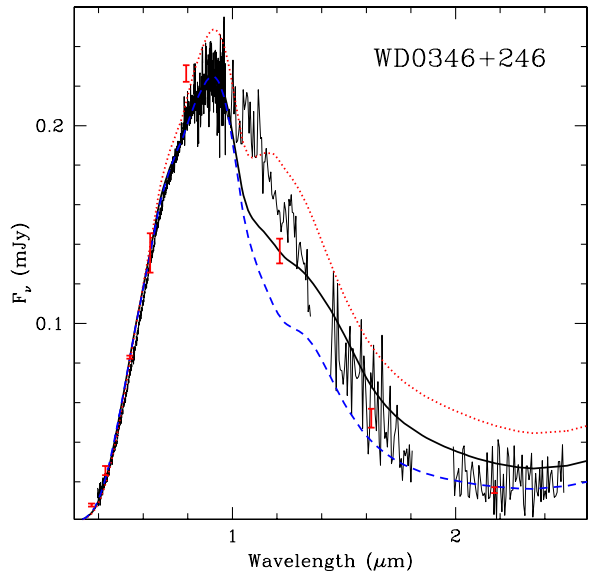}
\caption{The WD0346 SED \citep{oppenheimer01a} compared to the best-fit model with $T_{\rm eff}=3650$ K and He/H=0.43. Two other models with
$T_{\rm eff}=3500$ K (pure H atmosphere, dotted line) and $T_{\rm eff}=3600$ K (He/H=1, dashed line) are also shown.}
\end{figure}

In this paper we use the models with the aforementioned Ly-$\alpha$ opacity to derive the atmospheric parameters of WD0346.
The observed SED of WD0346 is best explained by a mixed atmosphere model with $T_{\rm eff}=3650$ K, $\log{g}=8.3$,
and He/H = 0.43, which is fairly consistent with the hydrogen-rich solution found by \citet{bergeron01}.  
Figure 3 shows this model plus two other models with different H/He compositions. Unlike J1102, the pure hydrogen atmosphere model cannot
reproduce the observed SED of WD0346 in the near-infrared. This is an indication that helium is present in the atmosphere and that through
collisions with hydrogen species it enhances the CIA opacity. The best-fit mixed H/He atmosphere model matches the optical portion of the
SED extremely well; the temperature of the star is constrained accurately by the optical data. However, the fit to the near-infrared data is
not perfect. As in J1102, the models predict a broad absorption feature around 1-1.2 $\mu$m that is not observed.
Regardless of these issues, WD0346 is clearly a very cool and old WD. Our best-fit model implies a mass of 0.77 $M_{\odot}$ and a cooling
age of $11.2^{+0.3}_{-1.6}$ Gyr.

\section{Discussion}

\subsection{Total Ages}

\citet{hall08} classify J1102 as either a pure-hydrogen atmosphere WD at $T_{\rm eff}=3830$ K (the most probable solution) or a mixed H/He atmosphere WD with H/He $\approx10^{-5}$
and $T_{\rm eff}=3500$ K. Thanks to our {\em Spitzer} photometry and parallax measurement, we are now able to constrain the atmospheric composition,
mass, and age of this WD. Our detailed model atmosphere analysis based on optical/infrared photometry and distance measurements for J1102 
demonstrate that it is a 3830 K, 0.62
$M_{\odot}$ star with a WD cooling age of $10.0^{+0.4}_{-1.1}$ Gyr. The initial-to-final mass relation for WDs indicate that the progenitor of J1102 was a
1.8-2.2 $M_{\odot}$ star \citep{catalan08,kalirai09,williams09} with a main-sequence lifetime of 0.6-1.1 Gyr \citep{marigo08}. Hence the total age of this object is
10.6-11.1 Gyr. Similary, WD0346 is a 3650 K, 0.77 $M_{\odot}$ star with a WD cooling age of $11.2^{+0.3}_{-1.6}$ Gyr. The progenitor star was a
3.1-3.3 $M_{\odot}$ main-sequence star with a main-sequence lifetime of 240-270 Myr \citep{marigo08}.
The uncertainty in main-sequence age is larger than the range given here. However, this uncertainty is insignificant
compared to the total age of the WD, which is 11.5 Gyr for WD0346. Both WD0346 and J1102 are significantly older than the coolest
disk WDs known \citep{leggett98,kilic10b}.

\subsection{Halo Membership}

At a Galactic latitude of $+64^{\circ}$, J1102 is 50 pc above the Galactic plane. Assuming zero radial velocity, it has
$U= 68 \pm 5 , V= -259 \pm 16$, and $W= 50 \pm 3$ km s$^{-1}$ with respect to the local standard of rest \citep{schonrich10}.
The unknown radial velocity mostly affects the $W$ velocity. For example, a change in the radial velocity from $-50$ to +50 km s$^{-1}$
corresponds to a change in $W$ from 5 to 95 km s$^{-1}$ and $V$ from $-261$ to $-256$ km s$^{-1}$.
J1102 lags behind the Galactic disk. Figure 4 plots the Galactic orbit of J1102 for the past and the next 1 Gyr, in a static
disk-halo-bulge potential \citep{kenyon08}. J1102 goes above and below the plane by as much as 5 kpc and its closest approach to the Galactic
center occurs at a distance of $R\approx$ 150 pc. Even though J1102 is only 33.7 pc away from us right now, it will travel as far away as 16
kpc within the next 1 Gyr. J1102 is clearly a halo WD.

At a Galactic latitude of $-23^{\circ}$, WD0346 is 9 pc above the plane. Assuming zero radial velocity, it has
$U= -2 \pm 7 , V= -148 \pm 22$, and $W= -55 \pm 9$ km s$^{-1}$ with respect to the local standard of 
rest\footnote{The $UVW$ velocities for WD0346 were miscalculated by \citet{kilic10a} due to a sign error in its proper motion.}.
The unknown radial velocity mostly affects the $U$ velocity. For example, a change in the radial velocity from $-50$ to +50 km s$^{-1}$
corresponds to a change in $U$ from +43 to $-47$ km s$^{-1}$. WD0346 also lags behind the Galactic disk.
WD0346 goes above and below the plane by as much as 2.1 kpc and its closest approach to the Galactic center occurs at a distance of $R\approx$ 2.2 kpc.
Even though WD0346 is only 28 pc away from us right now, it will also travel as far away as 16 kpc within the next 1 Gyr.
WD0346 is most likely a halo WD.

\begin{figure}
\includegraphics[width=3.2in,angle=0]{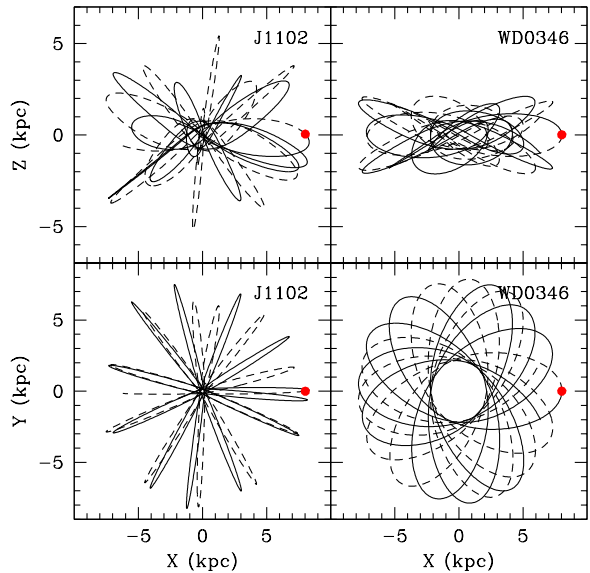}
\caption{The Galactic orbits of J1102 and WD0346 for the past 1 Gyr (solid lines) and the next 1 Gyr (dashed lines).
The current position of each WD is marked with a dot.}
\end{figure}

\subsection{The Oldest Stars in the Solar Neighborhood}

We now know of stars that are almost as old as the Universe \citep{hill02}. However, old stars with reliable age measurements are rare;
radioctive elements like thorium and uranium are detected only in a few cases, and even then the age uncertainties are $\sim$2-3 Gyr \citep{frebel07}.
Compared to main-sequence and giant stars, WDs provide
an advantage in age measurements, thanks to the relatively short main-sequence lifetimes of typical 0.6 $M_{\odot}$ WDs and the well understood
physics of the cooling WDs. \citet{sion09} present a volume limited sample of WDs within 20 pc of the Sun. \citet{giammichele12} show that
the coolest WDs in that sample are about 4200 K; significantly hotter than the two WDs discussed in this paper.

With estimated ages between 11 and 12 Gyr, J1102 and WD0346 are currently the oldest stars known in the solar neighborhood. The theoretical uncertainties
due to the unknown core composition, helium layer mass, crystallization, and phase separation are $\sim$1 Gyr for these ages \citep{montgomery99}.
The age estimates for these two WDs are remarkably similar to the oldest WDs found in the nearest globular clusters M4 and NGC 6397.
J1102 and WD0346 are about 100 times closer than these clusters. They provide an unprecedented opportunity to understand the model
uncertainties in cool WD atmospheres and put the globular cluster ages on a more secure footing.
Our analysis shows that the current WD atmosphere models including the Ly-$\alpha$ opacity in the blue
and the CIA in the infrared provide a reasonable match to the observations, though problems most likely associated with the CIA calculations remain.
Infrared data are essential for constraining the atmospheric composition of cool WDs (see Fig. 2 and 3). Hence, the temperature and age estimates based
on two filter {\em HST} optical data are likely uncertain by a few hundred degrees and several hundred Myr, respectively.

\section{Conclusions}

We present trigonotometric parallax and {\em Spitzer} mid-infrared photometric observations of the halo WD candidate J1102. We use these new data
along with optical and near-infrared data on J1102 to constrain the temperature, mass, and age of this WD. J1102 is an 11 Gyr old WD 33.7 pc away from
the Sun. We also revisit the model atmosphere analysis of another halo WD candidate, WD0346, using improved
model atmosphere calculations. WD0346 is an 11.5 Gyr old WD 28 pc away from the Sun. WD0346 and J1102 are currently the oldest stars known in the
solar neighborhood. Both stars display halo kinematics; they are just visiting our neighborhood at the moment.

WD0346 and J1102 are the best examples of serendipitious discoveries of nearby halo WDs. Current and future deep, wide-field surveys ought to find many more
old halo WDs. Such WDs remain to be discovered in Pan-STARRS, Palomar Transient Factory, the Large Synoptic Survey Telescope, and GAIA data. These discoveries
will provide independent age measurements and constrain the age and age range of the Galactic halo.

\section*{Acknowledgements}
MK is grateful to the Ohio State University Astronomy Department for large
amounts of time allocated on the MDM 2.4m Telescope. We thank Warren Brown
for useful discussions and Sandy Leggett for a detailed and constructive referee report.  
JRT gratefully acknowledges support from the NSF (awards AST-0708810 and AST-1008217).


\begin{thebibliography}{}
\bibitem[Bergeron(2001)]{bergeron01} Bergeron, P.\ 2001, \apj, 558, 369
\bibitem[Bergeron et al.(2005)]{bergeron05} Bergeron, P., Ruiz, M.~T., Hamuy, M., Leggett, S.~K., Currie, M.~J., Lajoie, C.-P., \& Dufour, P.\ 2005, \apj, 625, 838
\bibitem[Catal{\'a}n et al.(2008)]{catalan08} Catal{\'a}n, S., Isern, J., Garc{\'{\i}}a-Berro, E., \& Ribas, I.\ 2008, \mnras, 387, 1693 
\bibitem[Durant et al. (2012)]{DUR12} Durant, M., Kargaltsev, O. Pavlov, G. G., Kowalski, P. M., Posselt,B., van Kerkwijk, M. H., Kaplan, D. L. \ 2012, \apj, 746, 6
\bibitem[Fazio et al.(2004)]{fazio04} Fazio, G.~G., Hora, J.~L., Allen, L.~E., et al.\ 2004, \apjs, 154, 10
\bibitem[Fontaine et al.(2001)]{fontaine01} Fontaine, G., Brassard, P., \& Bergeron, P.\ 2001, \pasp, 113, 409
\bibitem[Frebel et al.(2007)]{frebel07} Frebel, A., Christlieb, N., Norris, J.~E., et al.\ 2007, \apjl, 660, L117 
\bibitem[Frommhold et al.(2010)]{frommhold10} Frommhold, L., Abel, M., Wang, F., Li, X., \& Hunt, K.~L.~C.\ 2010, AIP Conference Series, 1290, 219 
\bibitem[Gates et al.(2004)]{gates04} Gates, E. et al. 2004, \apj, 612, 129
\bibitem[Giammichele et al.(2012)]{giammichele12} Giammichele, N., Bergeron, P., \& Dufour, P.\ 2012, \apjs, in press, arXiv:1202.5581 
\bibitem[Hall et al.(2008)]{hall08} Hall, P.~B., Kowalski, P.~M., Harris, H.~C., Awal, A., Leggett, S.~K., Kilic, M., Anderson, S.~F., \& Gates, E.\ 2008, \aj, 136, 76
\bibitem[Hambly et al.(1997)]{hambly97} Hambly, N.~C., Smartt, S.~J., \& Hodgkin, S.~T.\ 1997, \apjl, 489, L157
\bibitem[Hambly et al.(1999)]{hambly99} Hambly, N.~C., Smartt, S.~J., Hodgkin, S.~T., et al.\ 1999, \mnras, 309, L33 
\bibitem[Hansen et al.(2004)]{hansen04} Hansen, B.~M.~S., et al.\ 2004, \apjs, 155, 551
\bibitem[Hansen et al.(2007)]{hansen07} Hansen, B.~M.~S., et al.\ 2007, \apj, 671, 380
\bibitem[Harris et al.(2006)]{harris06} Harris, H.~C., et al.\ 2006, \aj, 131, 571
\bibitem[Harris et al.(2008)]{harris08} Harris, H.~C., et al.\ 2008, \apj, 679, 697
\bibitem[Hill et al.(2002)]{hill02} Hill, V., Plez, B., Cayrel, R., et al.\ 2002, \aap, 387, 560
\bibitem[Kalirai et al.(2009)]{kalirai09} Kalirai, J.~S., Saul Davis, D., Richer, H.~B., et al.\ 2009, \apj, 705, 408 
\bibitem[Kalirai et al.(2012)]{kalirai12} Kalirai, J.~S., Richer, H.~B., Anderson, J., et al.\ 2012, \aj, 143, 11 
\bibitem[Kenyon et al.(2008)]{kenyon08} Kenyon, S.~J., Bromley, B.~C., Geller, M.~J., \& Brown, W.~R.\ 2008, \apj, 680, 312
\bibitem[Kilic et al.(2008a)]{kilic08} Kilic, M., Thorstensen, J.~R., \& Koester, D.\ 2008a, \apjl, 689, L45
\bibitem[Kilic et al.(2008b)]{K08} Kilic, M., Kowalski, P. M., Mullally, F., Reach, W. T. \& von Hippel, T. 2008b, \apj, 678, 1298
\bibitem[Kilic et al.(2009a)]{K09a} Kilic, M., Kowalski, P. M. \& von Hippel, T., 2009a, \aj, 138, 102
\bibitem[Kilic et al.(2009b)]{K09b} Kilic, M., Kowalski, P. M., Reach, W. T., von Hippel, T. 2009b, \apj, 696, 2094
\bibitem[Kilic et al.(2010a)]{kilic10a} Kilic, M., et al.\ 2010a, \apjl, 715, L21
\bibitem[Kilic et al.(2010b)]{kilic10b} Kilic, M., Leggett, S.~K., Tremblay, P.-E., et al.\ 2010b, \apjs, 190, 77 
\bibitem[Kowalski \& Saumon(2004)]{kowalski04} Kowalski, P. M. \& Saumon, D. 2004, \apj, 607, 970
\bibitem[Kowalski(2006a)]{kowalski06a} Kowalski, P. M. 2006a, \apj, 641,488
\bibitem[Kowalski(2006b)]{kowalski06c} Kowalski, P. M. 2006b, \apj, 651, 1120
\bibitem[Kowalski \& Saumon(2006)]{kowalski06b} Kowalski, P.~M., \& Saumon, D.\ 2006, \apjl, 651, L137
\bibitem[Kowalski et al.(2007)]{kowalski07} Kowalski, P. et al 2007, Phys. Rev. B. 2007, 76
\bibitem[Leggett et al.(1998)]{leggett98} Leggett, S.~K., Ruiz, M.~T., \& Bergeron, P.\ 1998, \apj, 497, 294
\bibitem[Liebert et al.(1988)]{liebert88} Liebert, J., Dahn, C. C., \& Monet, D. G. 1988, \apj, 332, 891
\bibitem[Marigo et al.(2008)]{marigo08} Marigo, P., Girardi, L., Bressan, A., Groenewegen, M.~A.~T., Silva, L., \& Granato, G.~L.\ 2008, \aap, 482, 883
\bibitem[Mestel(1952)]{mestel52} Mestel, L.\ 1952, \mnras, 112, 583 
\bibitem[Montgomery et al.(1999)]{montgomery99} Montgomery, M.~H., Klumpe, E.~W., Winget, D.~E., \& Wood, M.~A.\ 1999, \apj, 525, 482
\bibitem[Munn et al.(2004)]{munn04} Munn, J.~A., et al.\ 2004, \aj, 127, 3034
\bibitem[Oppenheimer et al.(2001a)]{oppenheimer01a} Oppenheimer, B.~R., Saumon, D., Hodgkin, S.~T., et al.\ 2001a, \apj, 550, 448 
\bibitem[Oppenheimer et al.(2001b)]{oppenheimer01b} Oppenheimer, B.~R., Hambly, N.~C., Digby, A.~P., Hodgkin, S.~T., \& Saumon, D.\ 2001b, Science, 292, 698 
\bibitem[Rohrmann et al. (2011)]{RAK11} Rohrmann, R. D., Althaus, L. G., \& Kepler, S. O.\ 2011, \mnras, 411, 781
\bibitem[Robin et al.(2003)]{robin03} Robin, A.~C., Reyl{\'e}, C., Derri{\`e}re, S., \& Picaud, S.\ 2003, \aap, 409, 523
\bibitem[Sch{\"o}nrich et al.(2010)]{schonrich10} Sch{\"o}nrich, R., Binney, J., \& Dehnen, W.\ 2010, \mnras, 403, 1829 
\bibitem[Sion et al.(2009)]{sion09} Sion, E.~M., Holberg, J.~B., Oswalt, T.~D., McCook, G.~P., \& Wasatonic, R.\ 2009, \aj, 138, 1681 
\bibitem[Thorstensen(2003)]{thor03} Thorstensen, J.~R.\ 2003, \aj, 126, 3017
\bibitem[Thorstensen et al.(2008)]{thor08} Thorstensen, J.~R., L\'epine, S. \& Shara, M.\ 2008, \aj, 136, 2107
\bibitem[Tonry et al.(2012)]{tonry12} Tonry, J.~L., Stubbs, C.~W., Kilic, M., et al.\ 2012, \apj, 745, 42
\bibitem[Williams et al.(2009)]{williams09} Williams, K.~A., Bolte, M., \& Koester, D.\ 2009, \apj, 693, 355 
\bibitem[Winget et al.(1987)]{winget87} Winget, D. E., Hansen, C. J., Liebert, J., Van Horn, H. M., Fontaine, G., Nather, R. E., Kepler, S. O., \& Lamb, D. Q. 1987, \apj, 315, L77
\bibitem[Wright et al.(2010)]{wright10} Wright, E.~L., Eisenhardt, P.~R.~M., Mainzer, A.~K., et al.\ 2010, \aj, 140, 1868 
\end{thebibliography}
\end{document}